\documentclass{article}

\usepackage{amsthm}
\usepackage{amsfonts}
\usepackage{amsmath}
\usepackage{amssymb}
\usepackage{algpseudocode}
\usepackage{algorithm}
\usepackage{array}
\usepackage{authblk}
\usepackage{bbm}
\usepackage{bm}
\usepackage{booktabs}
\usepackage{color}
\usepackage{enumitem}
\usepackage[font=small]{caption}
\usepackage{float}
\usepackage{helvet}
\usepackage{hyperref}
\usepackage{graphicx}
\usepackage{multirow}
\usepackage{mathtools}
\usepackage{url}
\usepackage{relsize}
\usepackage{setspace}
\usepackage{tabularx}

\usepackage[normalem]{ulem}
\newcommand{\revised}[1]{%
\ifx\highlightrevisions\undefined{#1}%
\else\textcolor{red}{#1}%
\fi}
\newcommand{\revisedtwo}[1]{%
\ifx\highlightrevisionstwo\undefined{#1}%
\else\textcolor{red}{#1}%
\fi}
\newcommand{\revisedthree}[1]{%
\ifx\highlightrevisionsthree\undefined{#1}%
\else\textcolor{red}{#1}%
\fi}
\newcommand{\rnum}[1]{%
\ifx\showreviewercommentnum\undefined%
\else{[\bf{R#1}] }%
\fi}
\newcommand{\del}[1]{%
\ifx\showerased\undefined%
\else{\sout{#1}}%
\fi}

\let\citeleft=(
\let\citeright=)

\begin{document}

\pdfinfo{
   /Author (Frank Ong, Xucheng Zhu, Joseph Y. Cheng, Kevin M. Johnson, Peder E. Z. Larson, Shreyas S. Vasanawala, and Michael Lustig)
   /Title (Extreme MRI: Large-Scale Volumetric Dynamic Imaging from Continuous Non-Gated Acquisitions)
}

\title{\vspace{-2cm}Extreme MRI: Large-Scale Volumetric Dynamic Imaging from Continuous Non-Gated Acquisitions}

\author[1]{Frank Ong}
\author[2]{Xucheng Zhu}
\author[3]{Joseph Y. Cheng}
\author[4, 5]{Kevin M. Johnson}
\author[6]{Peder E. Z. Larson}
\author[7]{Shreyas S. Vasanawala}
\author[8]{Michael Lustig}
    
\affil[1]{\small Electrical Engineering, Stanford University, CA. Work partly done at Electrical Engineering and Computer Sciences, University of California, Berkeley, CA.}
\affil[2]{UC Berkeley-UCSF Graduate Program in Bioengineering, University of California, San Francisco and University of California, Berkeley, CA}
\affil[3]{Work done at Radiology, Stanford University, CA. Currently at Apple Inc., CA.}
\affil[4]{Medical Physics, University of Wisconsin, Madison, WI.}
\affil[5]{Radiology, University of Wisconsin, Madison, WI.}
\affil[6]{Radiology and Biomedical Imaging, University of California, San Francisco, CA.}
\affil[7]{Radiology, Stanford University, CA.}
\affil[8]{Electrical Engineering and Computer Sciences, University of California, Berkeley, CA.}
\date{}
\maketitle

\noindent
\textit{Running head:} Extreme MRI

\noindent
\textit{Address correspondence to:} \\
  Michael Lustig \\
  506 Cory Hall \\
  University of California, Berkeley \\
  Berkeley, CA 94720 \\
  mlustig@eecs.berkeley.edu

\noindent
This work was supported by NIH R01EB009690, Bakar Family Fund, and GE Healthcare.

\noindent
Approximate word count: 247 (Abstract) 5000 (body)\\

\noindent
Submitted to \textit{Magnetic Resonance in Medicine} as a Full Paper.

\clearpage

\section*{Abstract}

\noindent
\textbf{Purpose}: To develop a framework to reconstruct large-scale volumetric dynamic MRI from rapid continuous and non-gated acquisitions, with applications to pulmonary and dynamic contrast enhanced (DCE) imaging.

  \noindent
  \textbf{Theory and Methods}: The problem considered here requires recovering hundred-gigabytes of dynamic volumetric image data from a few gigabytes of k-space data, acquired continuously over several minutes. This reconstruction is vastly under-determined, heavily stressing computing resources as well as memory management and storage. To overcome these challenges, we leverage intrinsic three dimensional (3D) trajectories, such as 3D radial and 3D cones, with ordering that incoherently cover time and k-space over the entire acquisition.  We then propose two innovations: (1) A compressed representation using multi-scale low rank matrix factorization that constrains the reconstruction problem, and reduces its memory footprint.  (2)  Stochastic optimization to reduce computation, improve memory locality, and minimize communications between threads and processors. We demonstrate the feasibility of the proposed method on DCE imaging acquired with a golden-angle ordered 3D cones trajectory and pulmonary imaging acquired with a bit-reversed ordered 3D radial trajectory. We compare it with ``soft-gated" dynamic reconstruction for DCE and respiratory resolved reconstruction for pulmonary imaging.

  \noindent
  \textbf{Results}: The proposed technique shows transient dynamics that are not seen in gating based methods. When applied to datasets with irregular,  or non-repetitive motions, the proposed method displays sharper image features.

  \noindent
  \textbf{Conclusion}: We demonstrated a method that can reconstruct massive 3D dynamic image series in the extreme undersampling and extreme computation setting.

\noindent
\textbf{Keywords}: Volumetric Dynamic MRI, Multiscale Low Rank, Stochastic Optimization, DCE-MRI, Pulmonary MRI

\clearpage

\section{Introduction}

Volumetric dynamic MRI has become an important component in a wide variety of applications, including pulmonary~\cite{johnson_optimized_2013, jiang_motion_2018}, flow~\cite{markl_4d_2012, cheng_comprehensive_2016}, and dynamic contrast enhanced (DCE) imaging~\cite{zhang_fast_2015}. Among its many advantages, 3D dynamic MRI enables high quality multiplanar and temporal reformatting, which can greatly enhance clinical interpretations. Recent advances in parallel imaging~\cite{pruessmann_sense:_1999, griswold_generalized_2002}, compressed sensing~\cite{lustig_sparse_2007} and data sorting techniques~\cite{feng_xd-grasp:_2016} have also substantially improved spatiotemporal resolution tradeoffs and motion robustness, which translate to many practical benefits in clinical settings~\cite{tariq_venous_2013, zhang_clinical_2014, zucker_free-breathing_2018}.

Because volumetric dynamic MRI reconstruction is inherently underdetermined, most existing methods~\cite{feng_xd-grasp:_2016, han_self-gated_2017, jiang_motion_2018, christodoulou_magnetic_2018} rely on gating and data binning techniques to reduce reconstruction undersampling rates. In particular, these techniques exploit the periodicity or repeatability of the underlying dynamics, such as cardiac and respiratory motion. k-space samples are acquired over many cycles and are then sorted according to their corresponding phase in the cycle. Data sorting can be accomplished by leveraging external navigator signals, such as respiratory bellow or navigators derived from the MR data itself (self-gating). One can then either reconstruct a single phase (hard gating) or a weighted combination of the phases (soft gating)~\cite{johnson_improved_2012, cheng_free-breathing_2015, forman_reduction_2015, zhang_fast_2015}. All phases can also be jointly reconstructed to produce motion resolved images. This can be extended to include multiple dimensions, such as the two-dimensional parametrization of cardiac and respiratory phases in XD-GRASP~\cite{feng_xd-grasp:_2016}. XD-flow~\cite{cheng_comprehensive_2017} further incorporates flow encoding. Recently, MR multitasking~\cite{christodoulou_magnetic_2018} considers multiparametric mappings and bins measurements in five dimensions.

The main drawback of gating and data binning methods is that periodic assumptions may not hold in practice. Incompatible dynamics include bulk movements, coughing and contrast enhancements. These transient dynamics can be lost in reconstruction or assigned incorrectly to one of the phases, causing image artifacts. Moreover, gating techniques require accurate estimations of the underlying respiratory signal, which can be challenging to obtain for patients with irregular motions. Finally, when integrated to applications like DCE, in which non-periodic dynamics are desired, these methods inherently limit the temporal resolution to be coarser than a respiratory cycle.

In this article, we aim to overcome limitations of periodicity constraints. In two-dimensional (2D) dynamic imaging, there are a wide range of works~\cite{riederer_mr_1988, holsinger_real-time_1990, kerr_real-time_1997, nayak_real-time_2004, uecker_real-time_2010, goud_real-time_2010, seiberlich_improved_2011} on reconstructing non-gated real-time dynamic MRI at a high spatial resolution. However, three-dimensional (3D) dynamic MRI reconstruction is vastly more underdetermined and demanding of computation and memory. While compressed sensing and low rank reconstruction techniques enable better tradeoffs between spatial and temporal resolution, state of the art methods~\cite{zhang_fast_2015, burdumy_one-second_2017, feng_grasppro:_2020} can only achieve temporal resolution on the order of seconds. Alternatively, spatial resolution is usually compromised to achieve higher frame-rates~\cite{chiew_k-t_2015, fu_highframerate_2017}. One reason for the limitations is that most existing techniques consider Cartesian or stack of 2D non-Cartesian trajectories, such as stack-of-stars or stack-of-spirals, which do not sample efficiently in at least one dimension. These sampling trajectories are often preferred partly because of their computational efficiency: the reconstruction problem can be broken down into 2D sub-problems that can easily fit into memory of local servers and deployed concurrently across processors.

Instead, we leverage intrinsic 3D non-Cartesian trajectories, such as 3D radial~\cite{johnson_optimized_2013} and 3D cones~\cite{gurney_design_2006}. These trajectories with pseudo-random orderings can efficiently cover k-space in a short time interval, making them ideal for high resolution volumetric dynamic imaging. Moreover, their undersampling artifacts are diffused in the image domain along all directions, which fit well to the compressed sensing framework. However, even with these efficient trajectories, the reconstruction problem is still heavily underdetermined, and significantly more computationally and memory expensive. In particular, the problem requires recovering images on the order of a hundred gigabytes (GB) from a few GBs of measurements, and \textbf{cannot be decoupled into smaller 2D sub-problems.}

We propose two innovations to overcome these reconstruction challenges: (1) a compressed representation using the multi-scale low rank matrix factorization (MSLR)~\cite{ong_beyond_2016} that  both constrains the reconstruction problem and reduces its memory footprint, and (2) stochastic optimization to reduce computation, improve memory locality and minimize communications between threads and processors.

\begin{figure}[!ht]
\begin{center}
  \includegraphics[width=\linewidth]{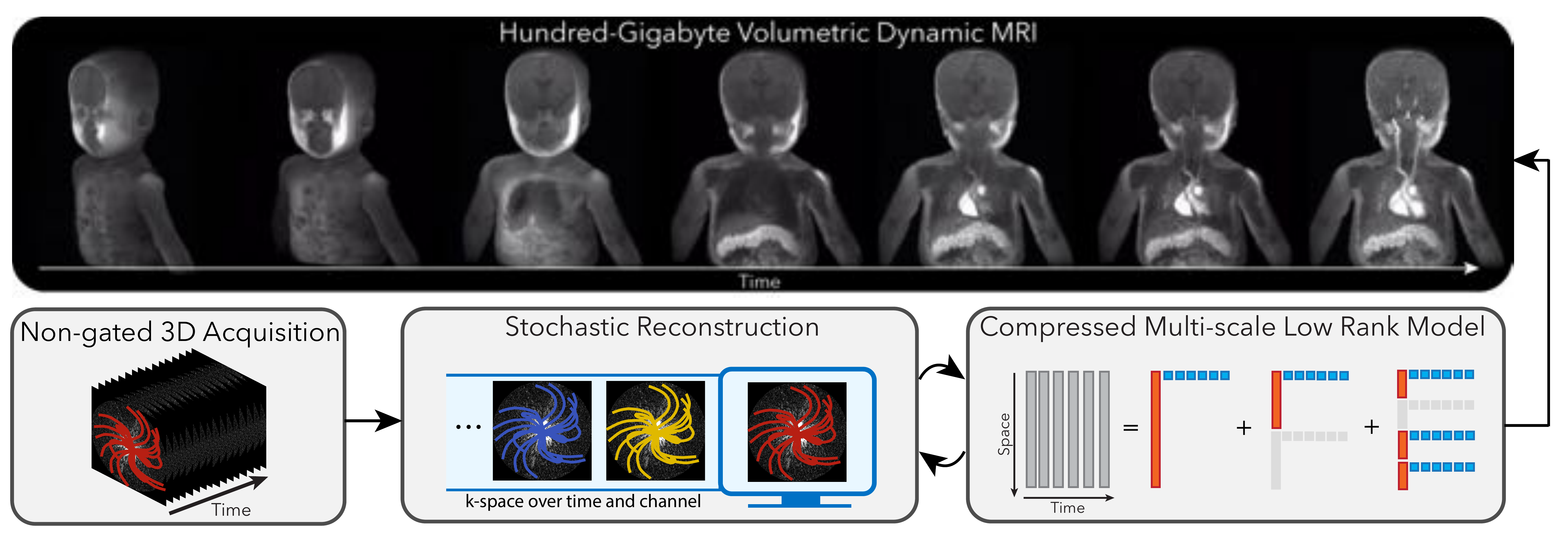}
\caption{Overview of the proposed method. We leverage intrinsic three dimensional (3D) trajectories, such as 3D radial and 3D cones, with ordering that incoherently cover time and k-space over the entire acquisition. We then propose two innovations: (1) a compressed representation using multi-scale low rank matrix factorization that both constrains the reconstruction problem and reduces its memory footprint.  (2)  Stochastic optimization to reduce computation, improve memory locality and minimize communications between threads and processors. This framework allows us to reconstruct large-scale volumetric dynamic MRI, which enables high quality reformatting as shown in Video~\ref{vid:dce1_3drender}.}
\label{fig:overview}
\end{center}
\end{figure}

MSLR was previously studied in~\cite{ong_beyond_2016} and generalizes low rank (LR)~\cite{liang_spatiotemporal_2007,pedersen_k-t_2009}, locally low rank (LLR)~\cite{trzasko_local_2011} and low rank + sparse (L+S)~\cite{otazo_low-rank_2015} matrix models. The representation considers all scales of correlations; local, global and sparse. Hence, it can obtain a more compact signal representation than other LR methods. In Section~\ref{sec:multiscale_low_rank}, we will combine the compact MSLR representation with the MRI acquisition model. A consequence of using this compressed representation is that the underlying dynamic sequence of images, which can require hundreds of GBs of storage, can be represented in mere few GBs. We propose an objective function that directly solves for the compressed representation. Such an approach makes it feasible to implement the reconstruction on local workstations. In the context of LR modeling, this is often referred to as the Burer-Monteiro factorization, which will be described in Section~\ref{sec:burer}.

To further reduce computation, we propose using stochastic gradient descent (SGD)~\cite{robbins_stochastic_1951} for reconstruction. SGD is commonly used nowadays in machine learning to efficiently optimize over large-scale datasets for training neural networks. Inspired by this use, we apply SGD to large-scale volumetric dynamic MRI reconstruction. SGD allows us to reduce the number of non-uniform fast Fourier transforms (NUFFT) from thousands-per-iteration to a single one per iteration. In particular, incorporating stochastic optimization for the proposed method cuts down the reconstruction time from weeks to hours, and will be described in more detail in Section~\ref{sec:stochastic}.

We demonstrate the feasibility of the proposed method in DCE imaging acquired with a golden-angle ordered 3D cones trajectory~\cite{gurney_design_2006} and lung imaging acquired with a bit-reversed ordered 3D ultra short time echo (UTE) radial trajectory~\cite{johnson_optimized_2013}. Our results show that the proposed technique, reconstructed at near millimeter spatial resolution and subsecond temporal resolution, can visualize certain transient dynamics that are lost in gating based reconstructions. 

\section{Theory}

\subsection{Multi-scale Low Rank (MSLR)}
\label{sec:multiscale_low_rank}

We begin by giving an overview of different LR models and motivate the use of MSLR for volumetric dynamic image reconstruction.

LR modeling~\cite{liang_spatiotemporal_2007, pedersen_k-t_2009} has been shown to be effective at representing static tissues, global contrast changes or smooth dynamics in many dynamic imaging applications~\cite{haldar_spatiotemporal_2010, goud_real-time_2010, lingala_accelerated_2011, zhao_image_2012, chiew_k-t_2015}.  \revised{\rnum{1.6}The model assumes that images over time are linear combinations of a small number of principal components. Equivalently, the spatiotemporal matrix, formed by stacking images as column vectors, has low rank.} Besides regularizing the reconstruction problem, explicit LR factorization also drastically reduces memory usage. This memory saving property was used in the many works~\cite{haldar_spatiotemporal_2010, goud_real-time_2010, lingala_accelerated_2011, zhao_image_2012, chiew_k-t_2015} for smaller data sizes and often for problems that can easily be decoupled into smaller 2D problems. In this article, the need for high compression rate becomes crucial as the problem cannot be separated into smaller 2D ones and is on the order of a hundred GB.

\revised{\rnum{2.4}On the other hand, LR modeling does not exploit spatial localities of the underlying dynamics. For example, blood vessel dynamics would require basis vectors of the size of a full image to be represented.} As a result, for problems requiring very low dimensional representations, spatially localized dynamics are often lost in LR reconstruction, as shown in Section~\ref{sec:results}. To mitigate this, LLR was proposed by Trzasko et al.~\cite{trzasko_local_2011} to exploit spatial locality by representing spatiotemporal matrices with block low rank matrices. While effective for spatially localized dynamics, LLR no longer exploit global correlation, which can lead to flickering artifacts in practice. Another direction to improve LR is the L+S matrix factorization technique~\cite{otazo_low-rank_2015}. L+S separately represents static background dynamics as a LR matrix and fast transient dynamics as a sparse matrix. However, unlike LR or LLR, L+S cannot be used for compression as non-zero locations of sparse components are not known before reconstruction. In particular, sparse components in L+S would require storing the entire image sequence in order to perform reconstruction, which defeats the purpose of using a compressed representation for memory savings. 

Here we adopt the MSLR representation that generalizes the above mentioned LR models to represent dynamics at multiple scales. In MSLR, the spatiotemporal matrix is modelled as a sum of block-wise low rank matrices with increasing scales of block sizes, as illustrated in Figure~\ref{fig:overview}. This allows us to combine the benefits of both LR and LLR, using the decomposition technique introduced in L+S.

Concretely, let $T$ be the number of frames, $N$ be the image size, and $J$ be the number of scales for MSLR. And for scale $j = 1, \ldots, J$, let $N_j$ be the block size, $B_j$ be the number of blocks and $K_j$ be the maximum block matrix rank. Then for each block $b = 1, \ldots, B_j$, let us define $\mathbf{L}_{jb} \in \mathbb{C}^{N_j \times K_j}$ be the block spatial bases, $\mathbf{R}_{jb} \in \mathbb{C}^{T \times K_j}$ be the block temporal bases, and $\mathbf{M}_{jb} \in \mathbb{C}^{N \times N_j}$ be a linear operator that embeds the input block matrix to the full image. The volumetric spatiotemporal matrix $\mathbf{X} \in \mathbb{C}^{N \times T}$ is represented as
\begin{equation}
  \mathbf{X} = \sum_{j = 1}^J \sum_{b = 1}^{B_j} \mathbf{M}_{jb} \mathbf{L}_{jb} \mathbf{R}_{jb}^H
  \label{eq:model_multiscale_low_rank_long}
\end{equation}

For notation simplicity, we consider stacked spatial and temporal bases $\mathbf{L}_j \in \mathbb{C}^{N_j B_j \times K_j}$  and $\mathbf{R}_j \in \mathbb{C}^{T B_j \times K_j}$ respectively and a linear operator $\mathcal{M}_j: \mathbb{C}^{N_j B_j \times T B_j} \rightarrow \mathbb{C}^{N \times T}$ that embeds the stacked input to an image such that
\begin{equation}
  \mathbf{X} = \sum_{j = 1}^J  \mathcal{M}_j \left(\mathbf{L}_j \mathbf{R}_j^H \right).
  \label{eq:model_multiscale_low_rank}
\end{equation}

\subsection{Dynamic MRI Forward Model}
\label{sec:forward_model}

We consider multi-channel MRI acquisition systems with $C$ coil channels and divide the overall scan into $T$ frames. We assume the underlying image for each frame is approximately static for fine enough temporal resolution. That is, we consider view sharing with short non-overlapping time windows. Let $M$ be the number of measurements for each frame, $\mathcal{A}: \mathbb{C}^{N \times T} \rightarrow \mathbb{C}^{MC \times T}$ be the overall sensing linear operator, which incorporates sensitivity maps and non-uniform Fourier sampling, and $\mathbf{W} \in \mathbb{C}^{MC \times T}$ represent the \revised{\rnum{2.7}complex} white Gaussian acquisition noise. Then, the k-space measurement matrix $\mathbf{Y} \in \mathbb{C}^{MC \times T}$ is given by
\begin{equation}
  \mathbf{Y} = \mathcal{A}(\mathbf{X}) + \mathbf{W}.
  \label{eq:model_mri}
\end{equation}

Since the number of measurements is vastly smaller than the total dynamic image size, the reconstruction problem is severely underdetermined. \revised{\rnum{2.1}\rnum{2.8}Compressed sensing MRI~\cite{lustig_sparse_2007} proposed a framework of using an incoherent sampling acquisition, a sparsifying signal representation, and a non-linear algorithm for underdetermined MRI reconstruction. While actual measurement systems deviate from theoretical assumptions in~\cite{candes_robust_2006, donoho_compressed_2006}, the framework provides practical guidance on sampling designs and motivated the use of low dimensional models.} Following this structure, we leverage intrinsic 3D non-Cartesian trajectories, such as 3D radial~\cite{johnson_optimized_2013} and 3D cones~\cite{gurney_design_2006}, with ordering that incoherently cover time and k-space over the entire acquisition. As for the underlying model, we use MSLR in the previous subsection for compactly representing the underlying volumetric image sequence. The overall forward model is then given by,
\begin{equation}
  \mathbf{Y} = \mathcal{A} \left(\sum_{j = 1}^J \mathcal{M}_j (\mathbf{L}_j \mathbf{R}_j^H) \right) + \mathbf{W}.
\end{equation}

\subsection{Memory efficient formulation using the Burer-Monteiro Factorization}
\label{sec:burer}

With the forward model, one way to impose low rank constraints is through convex relaxation by minimizing the nuclear norm~\cite{fazel_matrix_2002, recht_guaranteed_2010, candes_exact_2009}, or equivalently the sum of singular values. In particular, let $\| \mathbf{X} \|_*$ denote the nuclear norm of a matrix $\mathbf{X}$. Then for each scale $j$, let $\lambda_j$ be the regularization parameter and $\|\mathbf{X} \|_{(j)} = \sum_{b = 1}^{B_j} \| \mathbf{M}_{jb}^H \mathbf{X} \|_*$ denote the block-wise nuclear norm, the convex formulation considers the objective function,
\begin{equation}
  f(\mathbf{X}) = \frac{1}{2} \left\| \mathbf{Y} - \mathcal{A} \left(\sum_{j = 1}^J \mathcal{M}_j(\mathbf{X}_j) \right) \right\|_2^2
  + \sum_{j = 1}^J \lambda_i \| \mathbf{X}_i \|_{(j)}.
  \label{eq:cvx_objective_function}
\end{equation}

\revised{\rnum{2.1}When the measurement system is the identity matrix, theoretical properties of MSLR decomposition using the nuclear norm were studied in~\cite{ong_beyond_2016}. While this setting substantially deviates from our forward model, it provides practical guidance on selecting regularization parameters. In particular, under the identity setup and certain idealized incoherent conditions on the basis vectors, MSLR decomposition can be recovered when $\{\lambda_j\}_{j = 1}^J$ are chosen such that
\begin{equation}
    \lambda_j \propto \sqrt{N_j} + \sqrt{T} + \sqrt{2 \log B_j}.
    \label{eq:lambdas}
\end{equation}
This relationship is useful because we would only need to tune one scaling parameter instead of $J$ regularization parameters. In our experiments, we chose $\{\lambda_j\}_{j = 1}^J$ following Eq.~\ref{eq:lambdas}.}

However, a significant downside of the convex formulation is that it uses significant amount of memory. This is because full sized spatiotemporal matrices are stored even when they are extremely low rank. In particular, for an image size of $320 \times 320 \times 320$ and 500 frames, merely storing the image in complex single precision floats requires $125$ GBs! Applying iterative algorithms would require a few times more memory for work-space, which can approach terabytes. Local workstations, or even computing clusters, have difficulties handling such memory demand.

Rather than minimizing the convex problem in Equation~\eqref{eq:cvx_objective_function}, we consider the Burer-Monteiro factorization~\cite{burer_nonlinear_2003, recht_guaranteed_2010} to directly solve for the compressed representation. It considers the following non-convex transformation,
\begin{equation}
   \| \mathbf{X} \|_* = \min_{\mathbf{X} = \mathbf{L} \mathbf{R}^H} \frac{1}{2} (\| \mathbf{L} \|_F^2 + \| \mathbf{R} \|_F^2),
\end{equation}
and minimizes the following equivalent objective function:
\begin{equation}
  f(\mathbf{L}, \mathbf{R}) = \frac{1}{2} \left\| \mathbf{y} - \mathcal{A} \left(\sum_{j = 1}^J \mathcal{M}_j \left(\mathbf{L}_j \mathbf{R}_j^H \right) \right) \right\|_2^2  
  + \sum_{j = 1}^J \frac{\lambda_j}{2} \left( \| \mathbf{L}_j \|_F^2 + \| \mathbf{R}_j \|_F^2  \right).
  \label{eq:ncvx_objective_function}
\end{equation}

The primary benefit of this formulation is that it can use significantly less memory for low rank matrices. For a typical 3D volume, the variables can even fit the variables in graphic processing units (GPU). \revised{\rnum{2.1}However, the main drawback is that the problem becomes non-convex. Nonetheless, in practice, we observe that applying first-order gradient methods on the non-convex objective function reaches decent solutions. In the Discussion section, we point to recent theoretical results on the Burer-Monteiro factorization to motivate the non-convex formulation. However, we emphasize that there is currently no theoretical guarantees for the sensing system considered here.}

Before moving to the algorithm, we note that the proposed formulation can easily incorporate priors of the temporal dynamics, such as smoothness. For example, given a finite difference operator $\mathbf{D} \in \mathbb{C}^{T - 1 \times T}$, the above objective function can be modified to enforce temporal smoothness as follows:
\begin{equation}
  f(\mathbf{L}, \mathbf{R}) = \frac{1}{2} \left\| \mathbf{y} - \mathcal{A} \left(\sum_{j = 1}^J \mathcal{M}_j(\mathbf{L}_j \mathbf{R}_j^H)\right) \right\|_2^2  
  + \sum_{j = 1}^J \frac{\lambda_i}{2} \left( \| \mathbf{L}_i \|_F^2 + \| \mathbf{D} \mathbf{R}_j \|_F^2  \right).
  \label{eq:ncvx_objective_function_fd}
\end{equation}

\subsection{Stochastic Reconstruction}
\label{sec:stochastic}

Even with the explicit LR factorization, performing conventional iterative reconstruction is still prohibitively slow, requiring computing many NUFFTs per iteration, one for each time frame and channel. To address this, we propose to apply stochastic optimization techniques to accelerate the reconstruction, which update variables in each iteration with random subsets of measurements.

The idea of using subsets of measurements for iterative updates has been proposed in the general medical imaging community. Examples include ordered subset algorithms for positron emission tomography reconstruction~\cite{hudson_accelerated_1994, erdogan_ordered_1999} and the algebraic reconstruction technique for computed tomography~\cite{gordon_algebraic_1970}. \revised{\rnum{2.13}In MRI, Muckley et al.~\cite{muckley_accelerating_2014} applied incremental gradient methods to parallel imaging
reconstruction.} A recent work~\cite{mardani_tracking_2016} also proposed to use stochastic optimization for  dynamic MRI reconstruction, but only demonstrated on gated k-space datasets. Our contribution is to show how stochastic optimization can transform an almost computationally infeasible reconstruction to a practical one. 

In particular, let us split the objective function into separate frames and coils,
\begin{equation}
  f_{tc}(\mathbf{L}, \mathbf{R}) = \frac{1}{2} \left\| \mathbf{Y}_{tc} - \mathcal{A}_{tc} \left(\sum_{j = 1}^J \mathcal{M}_j \left(\mathbf{L}_j \mathbf{R}_{jt}^H \right) \right) \right\|_2^2  
  + \sum_{j = 1}^J \frac{\lambda_j}{2} \left( \frac{1}{TC}\left \| \mathbf{L}_j \right\|_F^2 +  \frac{1}{C} \left\| \mathbf{D} \mathbf{R}_{j} \right\|_F^2 \right)
  \label{eq:ncvx_objective_function_tc},
\end{equation}
for $t = 1, \ldots, T$ and $c = 1, \ldots C$, then $f(\mathbf{L}, \mathbf{R}) = \sum_{t = 1}^T \sum_{c = 1}^C f_{tc}(\mathbf{L}, \mathbf{R})$.

In each iteration, SGD randomly selects a frame $t$ and a coil $c$, and performs the following updates:
\begin{equation}
\begin{aligned}
    \mathbf{L} &= \mathbf{L} - \alpha TC \nabla_{\mathbf{L}} f_{tc}(\mathbf{L}, \mathbf{R})\\
    \mathbf{R}_t &= \mathbf{R}_t - \alpha C \nabla_{\mathbf{R}_t} f_{tc}(\mathbf{L}, \mathbf{R}),
  \label{eq:update}
\end{aligned}
\end{equation}
where $\alpha$ is a step-size parameter. In the Methods section, we describe more in detail how to select $\alpha$.

With SGD, the number of NUFFTs per iteration is drastically reduced to 1. Moreover, SGD can easily support parallel processing by performing mini-batch updates. Given $G$ parallel processors, such as multiple GPUs, each processor $g$ can choose a different frame $t$ and a different coil $c$, calculate the gradients  $\nabla_{\mathbf{L}} f_{tc}(\mathbf{L}, \mathbf{R}_t)$ and $\nabla_{\mathbf{R}_t} f_{tc}(\mathbf{L}, \mathbf{R}_t)$, and additively synchronize afterwards. 

Concretely, in each iteration, SGD randomly selects an index set $\mathcal{I} = \{(t_1, c_1), (t_2, c_2), \ldots, (t_G, c_G)\}$. Then, SGD performs
\begin{equation}
  \begin{aligned}
    \mathbf{L} &= \mathbf{L} - \frac{\alpha TC}{G} \sum_{(t, c) \in \mathcal{I}} \nabla_{\mathbf{L}} f_{tc}(\mathbf{L}, \mathbf{R})\\
    \mathbf{R}_t &= \mathbf{R}_t - \frac{\alpha C}{G} \sum_{c: (t, c) \in \mathcal{I}} \nabla_{\mathbf{R}_t} f_{tc}(\mathbf{L}, \mathbf{R})~~\text{for all}~t~\text{such that}~(t, c) \in \mathcal{I},
  \label{eq:minibatch_update}
\end{aligned}
\end{equation}
where only the summations are synchronized across the processors, thus minimizing expensive communications.

\section{Methods}
\label{sec:methods}

We demonstrated the feasibility of the proposed method on DCE datasets acquired with a golden-angle ordered 3D cones trajectory~\cite{gurney_design_2006} and lung datasets acquired with a bit-reversed ordered 3D UTE radial trajectory~\cite{johnson_optimized_2013}. In general, our experiments aim to show that the proposed techniques enables large-scale volumetric dynamic MRI reconstruction. On the other hand, we emphasize that targeted reconstruction resolution does not necessarily translate to the true apparent resolution. Dynamics can be blurred and features can be lost. Hence, another purpose of our comparisons is to see what additional dynamics can be seen in the proposed reconstruction and what artifacts arise.

We first describe implementation details used in both applications. The proposed reconstruction was implemented in Python using the package SigPy~\cite{ong_sigpy:_2019} on workstations with two Intel Xeon Gold CPUs and four Titan Xp GPUs. All operations, except loading the data and splitting into frames, were performed on the GPUs.

The number of scales in MSLR was determined by the GPU memory constraint. \revised{\rnum{1.1}In particular, we used three scales with block widths 32, 64 and 128 for small datasets (the first two DCE datasets and the first lung dataset) and block widths 64, 128 and 256 for large datasets (the third DCE dataset and the second lung dataset).} \revised{\rnum{2.2}Because MSLR, like orthogonal wavelet transforms, is not shift invariant, blocking artifacts can arise in practice.} To prevent this, each block is overlapped by half the block size in all spatial dimensions. Each voxel in the end was represented by eight basis vectors in each scale, including the overlapped bases.

To ensure the same parameters provide similar performance for different datasets, the forward operators and k-space datasets were normalized appropriately. In particular, the NUFFT operators were normalized by the maximum singular value of the NUFFT operator of the first frame, which was estimated using ten iterations of power method. The k-space data was scaled by $\| \mathbf{x}_\text{grid}\|_2 / \sqrt{T}$, where $\mathbf{x}_\text{grid}$ is the \revised{\rnum{2.14}time-averaged} density compensated gridding reconstructed image. We found that this normalization scheme made regularization parameters less sensitive to dataset scaling and the number of frames. \revised{\rnum{2.7}k-space data were pre-whitened using additional noise measurements collected with each scan.} \revised{\rnum{1.1}Sensitivity maps were estimated using ESPIRiT~\cite{uecker_espirit-eigenvalue_2014} on time-averaged k-space datasets.}

The spatial bases $\{\mathbf{L}_i\}_i$ and temporal bases $\{\mathbf{R}_i\}_i$ were initialized as complex white Gaussian noise vectors with unit norm. To choose the step-size $\alpha$, we first set it to 1, and run the iteration updates. If the iteration diverged with the gradient norm reaching the numerical limit, the reconstruction was restarted and the step-size was halved. In practice, very few iterations were needed to detect divergence, as SGD is known to diverge exponentially~\cite{bach_non-asymptotic_2011} when the step-size is too large. 

The regularization parameters were chosen to be $\lambda_j = \lambda (\sqrt{N_j} + \sqrt{T} + \sqrt{2 \log B_j})$, with $\lambda$ being a tuned parameter. Due to the small image size of the second DCE dataset described below, we performed parameter search on it for a 20-frame reconstruction and used the same parameters for all other datasets. In particular, $\lambda = 10^{-4}$ results in the best subjective tradeoff between noise-level and spatiotemporal blurring. Supporting Information Figure S1 shows an example of the tradeoffs. Following common practices in machine learning, SGD selects time frames and coils sequentially from shuffled instances of $\{1, \ldots, T\} \times \{1, \ldots C\}$. Objective values over iterations were also recorded to set the number of epochs. The number of epochs was 60, which means SGD goes through the entire k-space dataset 60 times, resulting in a total of $60 TC / G$ SGD update steps.

We compared the proposed reconstruction with gating based methods, which required respiratory motion signals. We estimated them from filtered k-space center signals. There are two reasons for using k-space DC based navigators: (1) it is self-gated, which avoids drifting and asynchronous issues associated with respiratory bellows, and (2) completely automated, which does not need manual selections of region of interests as in image navigators. Symmetric extension was performed on the k-space center before filtering to prevent edge effects. The filter was optimized in the least squares sense to have a pass band between 0.1 Hz and 1 Hz and stop band everywhere else with a transition width of 0.05 Hz. The filtered signal was then robustly normalized by subtracting the median, and dividing by the median absolute deviation.

When the patient scans were performed, the prescribed field-of-views (FOV) were all smaller than the patient bodies, often with patient's arms outside the FOV. For iterative reconstruction, small FOVs result in artifacts from model mismatch. \revised{\rnum{1.7}Instead of using the prescribed FOV, we estimated the matrix size from a low resolution time-averaged image with a large reconstruction FOV. In particular, the image was reconstructed at twice the prescribed FOV with a gridding reconstruction using the first 100 samples along all readouts. We then applied a threshold of 0.1 of its maximum amplitude to estimate the FOV.}

\subsection{DCE with Golden-angle Ordered 3D Cones Trajectory}

We applied the proposed reconstruction on three DCE datasets from pediatric patients to qualitatively evaluate its performance. All datasets were acquired on GE 3T scanners using a spoiled gradient echo (SPGR) sequence and a golden-angle ordered 3D cones trajectory. The first dataset is a chest scan with regular respiratory motion and little bulk motion. The second dataset is a chest scan with several large bulk movements throughout the scan. The third dataset is an abdominal scan with regular respiratory motion. Signal intensity curves were plotted at voxels at the left and right ventricles, and aorta for the first two datasets, and at the cortex, medulla, and aorta for the third dataset.

On the first dataset, we performed reconstructions with LR, LLR, and MSLR of 500 frames to evaluate how different LR model affects the reconstruction quality. Matrix ranks for LR and LLR were chosen so that they have similar number of parameters as MSLR. For all datasets, we compared the proposed reconstruction method of 500 frames to soft-gated dynamic reconstructions of 20 frames. The soft-gated dynamic reconstruction was similar to Zhang et al.~\cite{zhang_fast_2015} except with a MSLR regularization instead of LLR. Soft-gating weights were computed from the respiratory motion signal mentioned above with a threshold chosen from the 10th percentile of the signal, and a decay parameter of 1.

The first DCE dataset was acquired with a 16-channel coil array, scan time of 4 minutes 40 seconds, TE=0.1 ms, TR=5.8 ms, flip angle=14 degrees, and bandwidth=125 kHz. The number of readout points was 624, and the number of interleaves was 48129. The spatial resolution was reconstructed at $1 \times 1 \times 2.8$ mm$^3$, and the matrix size was $323 \times 186 \times 332$. The targeted reconstruction temporal resolution for the proposed method was 580 ms, and for the soft-gated reconstruction was 14.5 s.

The second DCE dataset was acquired with a 12-channel coil array, scan time of 5 minutes 11 seconds, TE=0.1 ms, TR=7.4 ms, flip angle 15 degrees, bandwidth=125 kHz. The number of readout points was 711, and the number of interleaves was 41861. The spatial resolution was reconstructed at $1 \times 1 \times 1.8$ mm$^3$, and the matrix size was $360 \times 156 \times 126$. The targeted reconstruction temporal resolution for the proposed method was 614 ms, and for the soft-gated reconstruction was 15.49 s.

The third DCE dataset was acquired with a 32-channel coil array, scan time of 3 minutes 53 seconds, TE=0.1 ms, TR=8.3 ms, flip angle 14 degrees, bandwidth=125 kHz. The number of readout points was 1007 and the number of interleaves was 28083. The spatial resolution was reconstructed at $1 \times 1 \times 1.8$ mm$^3$, and the matrix size was $438 \times 276 \times 138$. The targeted reconstruction temporal resolution for the proposed method was 467 ms, and for the soft-gated reconstruction was 11.68 s.

\subsection{Pulmonary Imaging with Bit-reversed Ordered 3D UTE Radial Trajectory}

We applied the proposed reconstruction on two lung datasets from adult patients to qualitatively evaluate its performance. All datasets were acquired on GE 3T Discovery MR750 clinical scanners (GE Healthcare, Waukesha, WI) using an optimized UTE sequence~\cite{johnson_optimized_2013} and a bit-reversed ordered 3D radial trajectory. The first dataset has regular respiratory motion and little bulk motion. The second dataset has abrupt bulk motions such as coughing throughout the scan. 

We compared the proposed method with respiratory resolved reconstruction to five motion states. The proposed technique was reconstructed with 500 frames, whereas for the third lung dataset. The respiratory resolved reconstruction was performed with total variation regularization along the motion states, which is similar to Jiang et al.~\cite{jiang_motion_2018} and Feng et al.~\cite{feng_simultaneous_2019}. To exclude data corrupted by bulk motion, only k-space data with the respiratory signal between the 10th and 90th percentiles were used. They were then sorted into five equally sized bins, each representing a motion state.

More specifically, the first lung dataset was acquired with an 8-channel coil array, an overall scan time of 4 minutes 40 seconds, TE=80 $\mu$s, TR=2.81 ms, flip angle=4 degrees, and sampling bandwidth=250 kHz. The number of readout points was 523, and the number of interleaves was 99680. The targeted reconstruction spatial resolution was $1.25 \times 1.25 \times 1.25$ mm$^3$, and the matrix size was $500 \times 250 \times 241$. The targeted reconstruction temporal resolution was 560 ms. 

The second lung dataset was acquired with an 8-channel coil array, an overall scan time of 4 minutes 18 seconds, TE=80 $\mu$s, TR=3.48 ms, flip angle 4 degrees, and sampling bandwidth=250 kHz. The number of readout points was 654, and the number of interleaves was 75768. The targeted reconstruction spatial resolution was $1.25 \times 1.25 \times 1.25$ mm$^3$, and the matrix size was $408 \times 183 \times 379$. The targeted reconstruction temporal resolution was 515 ms. 


\section{Results}
\label{sec:results}

In the spirit of reproducible research, we provide a software package in Python to reproduce the results described in this chapter. The software package can be downloaded from:
\begin{center}
  \url{https://github.com/mikgroup/extreme_mri.git}
\end{center}

\subsection{DCE with Golden-angle Ordered 3D Cones Trajectory}

\subsubsection{First DCE dataset}
\begin{figure}[!ht]
\begin{center}
  \includegraphics[width=0.8\linewidth]{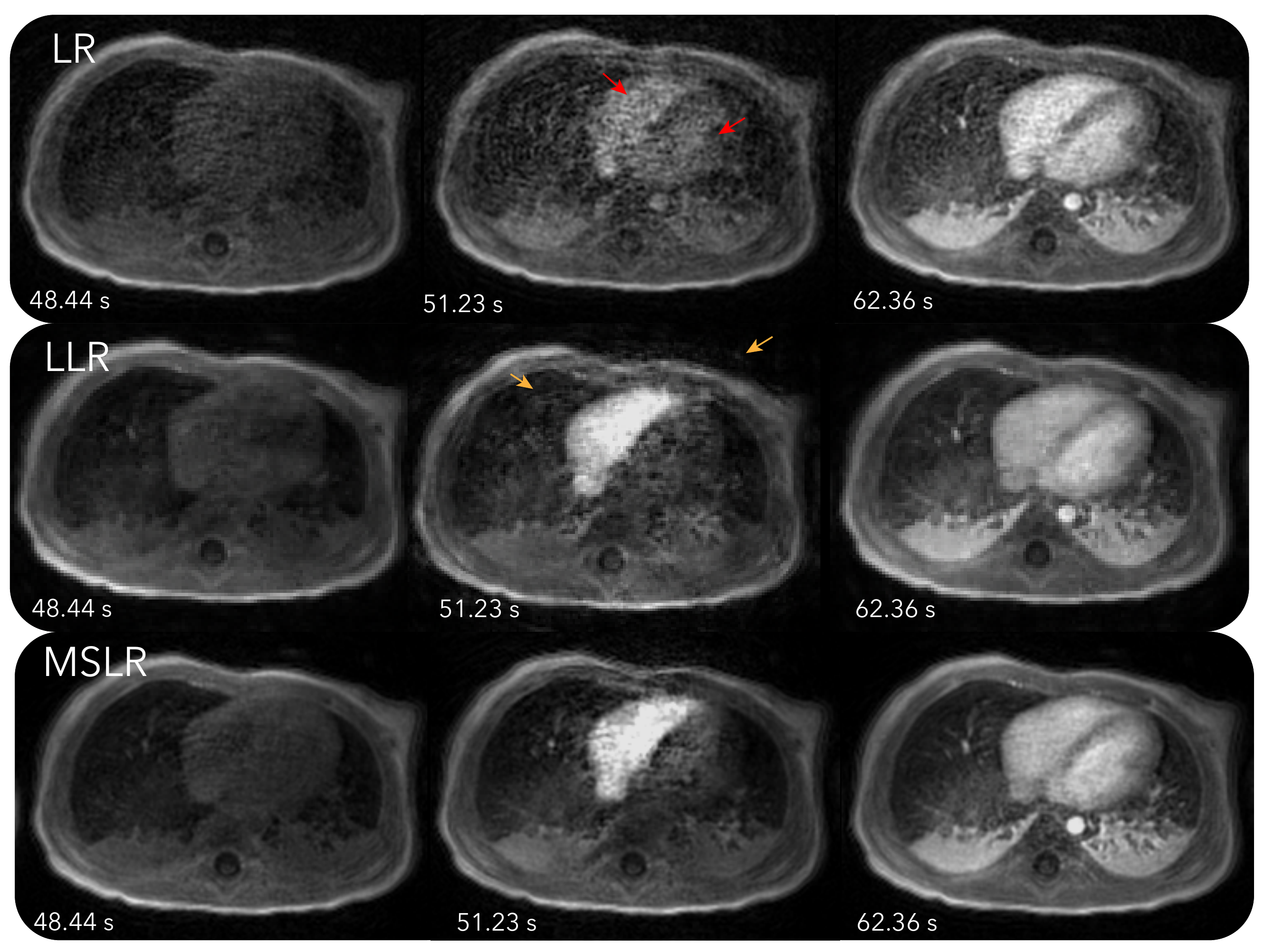}
\caption{Reconstructions with LR, LLR and MSLR on the first DCE dataset. Dynamics can be seen more clearly in Video~\ref{vid:dce1_lr}, \ref{vid:dce1_llr}, and \ref{vid:dce1_extreme}. As pointed by the red arrows, reconstruction with LR shows unrealistic dynamics, with contrast enhancing in both left and right ventricles at the same time. The other reconstructions show contrast enhancing first in the right ventricle then the left ventricle, which is physiologically correct. On the other hand, LLR displays more flickering temporal artifacts, which can be seen more visibly from Video~\ref{vid:dce1_llr}. These artifacts are also pointed out by the orange arrows in figure. MSLR achieves a balance between representing contrast enhancement dynamics and reducing artifacts.}
\label{fig:lr_compare}
\end{center}
\end{figure}

Figure~\ref{fig:lr_compare} and Video~\ref{vid:dce1_lr},~\ref{vid:dce1_llr}, and~\ref{vid:dce1_extreme} show reconstruction results with LR, LLR and MSLR. As pointed by the red arrows, reconstruction with LR shows unrealistic dynamics, with contrast enhancing in both left and right ventricles at the same time. The other reconstructions show contrast enhancing first in the right ventricle then the left ventricle, which is physiologically correct. On the other hand, LLR displays more flickering temporal artifacts, which can be seen more visibly from Video~\ref{vid:dce1_llr}. These artifacts are also pointed out by the orange arrows in figure. MSLR achieves a balance between representing contrast enhancement dynamics and reducing artifacts.

\begin{figure}[!ht]
\begin{center}
  \includegraphics[width=0.8\linewidth]{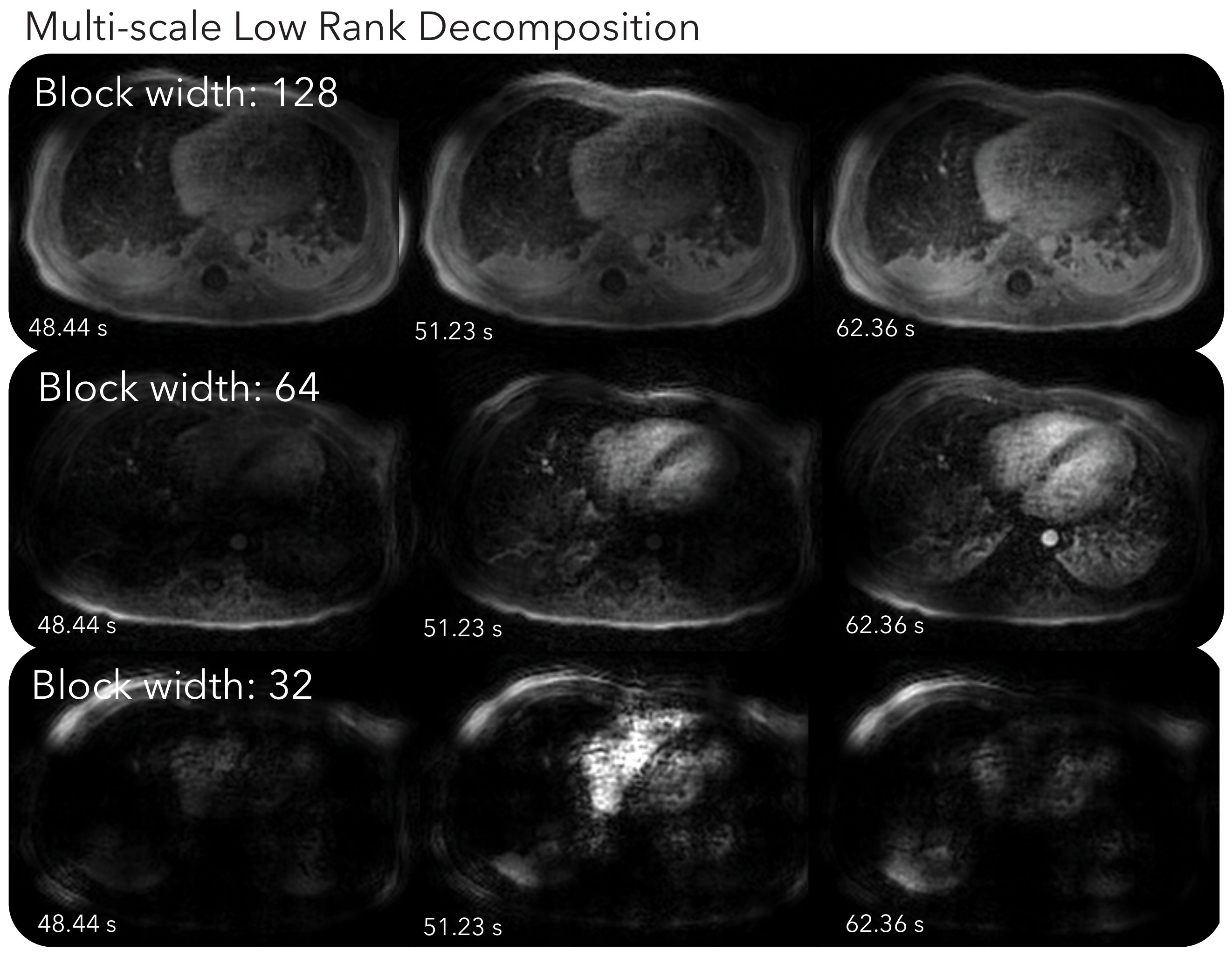}
\caption{MSLR decomposition of the first DCE dataset. Dynamics can be seen more clearly in Video~\ref{vid:dce1_mslr_decom}. The scale with the 128$^3$-sized blocks mostly shows static background tissues. The scale with the 64$^3$-sized blocks depicts mostly contrast enhancements in the heart and aorta, and respiratory motion. The scale with 32$^3$-sized blocks displays spatially localized dynamics, such as those in the right ventricle, and oscillates more over time than other scales.}
\label{fig:decom}
\end{center}
\end{figure}

Figure~\ref{fig:decom} and Supporting Information Video~\ref{vid:dce1_mslr_decom} show the corresponding MSLR decomposition. The scale with the 128$^3$-sized blocks mostly shows static background tissues. The scale with the 64$^3$-sized blocks depicts mostly contrast enhancements in the heart and aorta, and respiratory motion. The scale with 32$^3$-sized blocks displays spatially localized dynamics, such as those in the right ventricle, and oscillates more over time than other scales.

\begin{figure}[!ht]
\begin{center}
  \includegraphics[width=\linewidth]{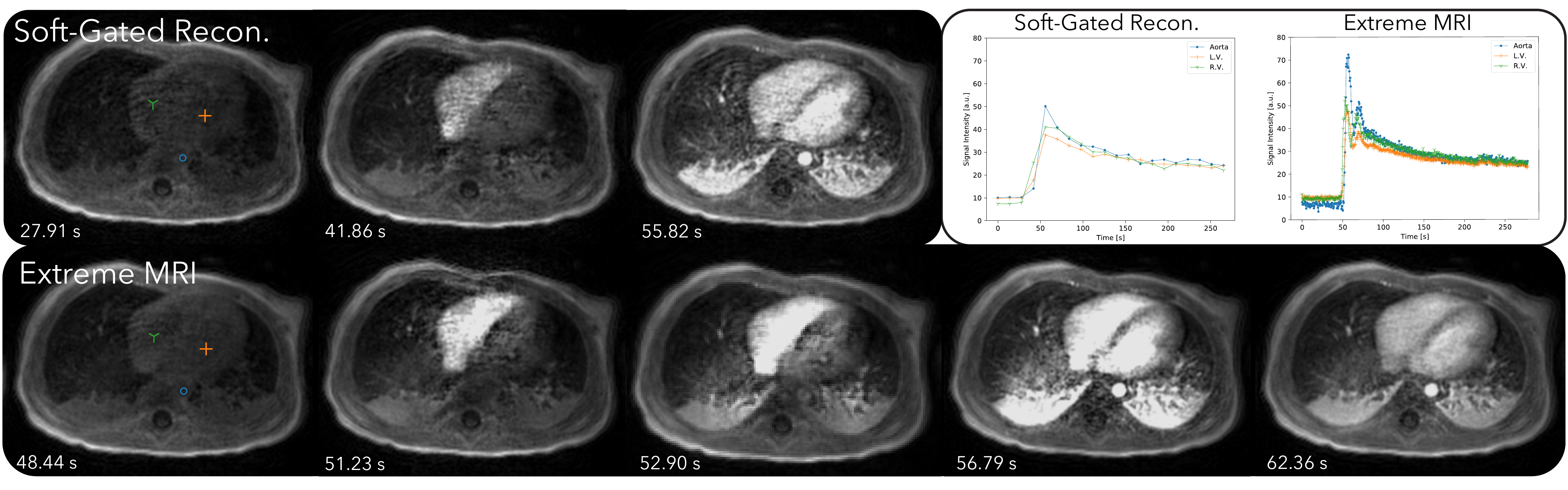}
\caption{Comparison of the proposed method with the soft-gated reconstruction of the first DCE dataset. Dynamics can be seen more clearly in Supporting Information Video~\ref{vid:dce1_extreme} and~\ref{vid:dce1_softgated}. Regular respiratory motion can be observed in the proposed reconstruction in Supporting Information Video~\ref{vid:dce1_extreme}. Contrast enhancements can be seen starting from the right ventricle, to the lung, then the left ventricle, and to the aorta, whereas the soft-gated reconstruction merges together the temporal changes of the lung, the left ventricle and the aorta. The signal intensity plots for the proposed reconstruction also show much higher peaks when contrast is injected, which is physiologically accurate. In comparison, the soft-gated reconstruction signal curves appear to be smoothed.}
\label{fig:dce1}
\end{center}
\end{figure}

Figure~\ref{fig:dce1} and Supporting Information Video~\ref{vid:dce1_extreme} and~\ref{vid:dce1_softgated} compare the proposed method with the soft-gated reconstruction. Regular respiratory motion can be observed in the proposed reconstruction in Video~\ref{vid:dce1_extreme}. \revised{\rnum{1.2}Slight movements of the subject's left arm can be seen from the video right before the contrast enhancements and toward the end of the scan. This can also be visualized in the 3D rendered Supporting Information Video~\ref{vid:dce1_3drender}.} Contrast enhancements can be seen starting from the right ventricle, to the lung, then the left ventricle, and to the aorta, whereas the soft-gated reconstruction merges together the temporal changes of the lung, the left ventricle and the aorta. The signal intensity plots for the proposed reconstruction also show much higher peaks when contrast is injected, which is physiologically accurate. In comparison, the soft-gated reconstruction signal curves appear to be smoothed.

An instance of the proposed method takes about 14 hours and the soft-gated reconstruction takes about 40 minutes. The resulting image using the MSLR representation takes 1.7 GBs to store.

\subsubsection{Second DCE dataset}

\begin{figure}[!ht]
\begin{center}
  \includegraphics[width=\linewidth]{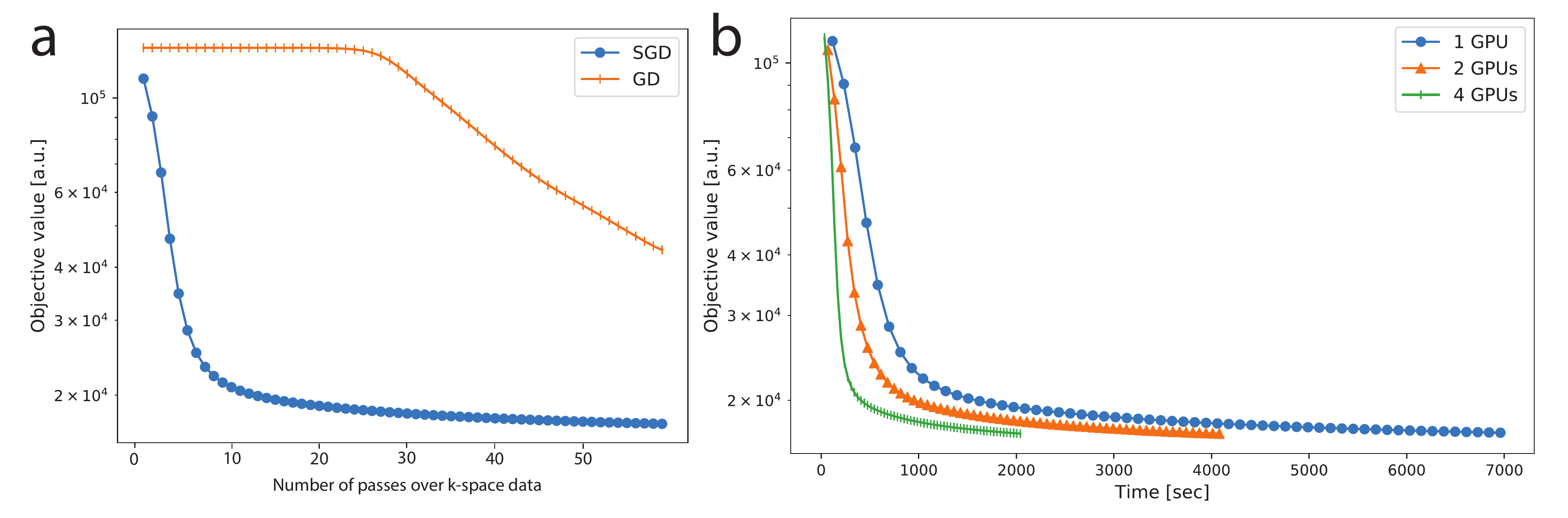}
\caption{Convergence plots for the second DCE dataset reconstructed with 20 frames. Each marker indicates a pass over the entire k-space dataset. \textbf{a)} Comparison between SGD with one GPU to GD. After 60 passes of k-space data, SGD is mostly converged, whereas GD only attains the objective value of about the fourth pass for SGD.  \textbf{b)} Comparison of SGD convergence rates with multiple GPUs. The convergence rates for different GPUs are similar. The convergence rates are similar in terms of number of passes. As for overall computation time, there is a small overhead going from one GPU to two GPUs, resulting in a 1.7x speedup instead of the ideal 2x speedup. We conjecture that this is due to data transferring between GPUs. From two to four GPUs, there is not much additional overhead, resulting in a speedup of almost 2x. The overall speedup from one to four GPUs is about 3.4.}
\label{fig:convergence}
\end{center}
\end{figure}

Convergence plots for the second DCE dataset reconstructed with 20 frames are shown in Figure~\ref{fig:convergence}a and b. Each marker indicates a pass through the entire k-space dataset. Figure~\ref{fig:convergence}a compares SGD with one GPU to GD. After 60 passes of k-space data, SGD is mostly converged, whereas GD only attains the objective value of about the fourth pass for SGD. Figure~\ref{fig:convergence}b compares the convergence of SGD with multiple GPUs. The convergence rates are similar in terms of number of passes. As for overall computation time, there is a small overhead going from one GPU to two GPUs, resulting in a 1.7x speedup instead of the ideal 2x speedup. We conjecture that this is due to data transferring between GPUs. From two to four GPUs, there is not much additional overhead, resulting in a speedup of almost 2x. The overall speedup from one to four GPUs is about 3.4.

\begin{figure}[!ht]
\begin{center}
  \includegraphics[width=\linewidth]{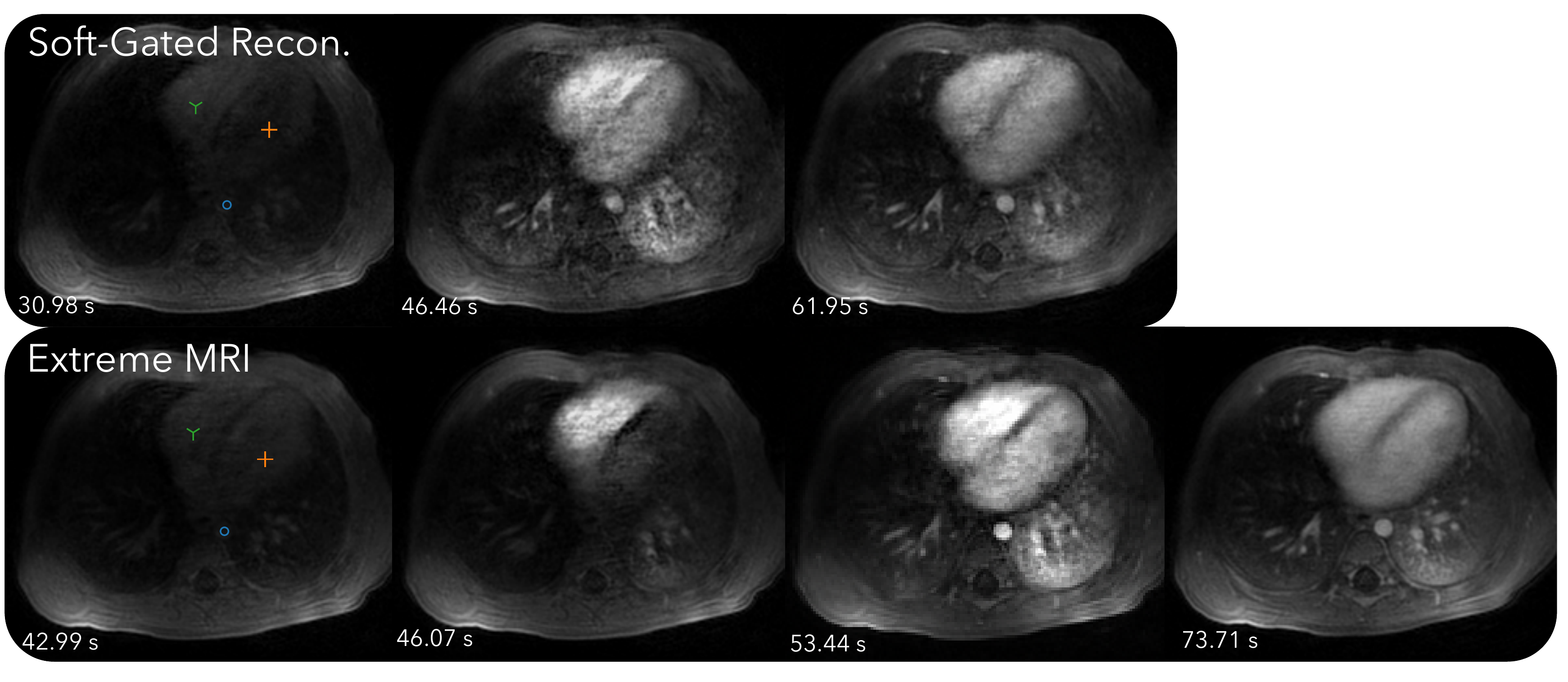}
\caption{Comparison of the proposed method with the soft-gated reconstruction of the second DCE dataset. Dynamics can be seen more clearly in Supporting Information Video~\ref{vid:dce2_extreme} and~\ref{vid:dce2_softgated}. From Supporting Information Video~\ref{vid:dce2_extreme}, the proposed reconstruction shows regular respiratory motion in the beginning, but after contrast injection, breathing becomes more rapid and the patient body shifts to the right seven times. While the image quality during these bulk movements degrades, it improves as soon as the patient body returns to the original position. Similar to the first DCE dataset, distinct phases of contrast enhancement to different organs can be seen, whereas the soft-gated reconstruction merges all dynamics into one frame, including the bulk motion.}
\label{fig:dce2}
\end{center}
\end{figure}

\begin{figure}[!ht]
\begin{center}
  \includegraphics[width=\linewidth]{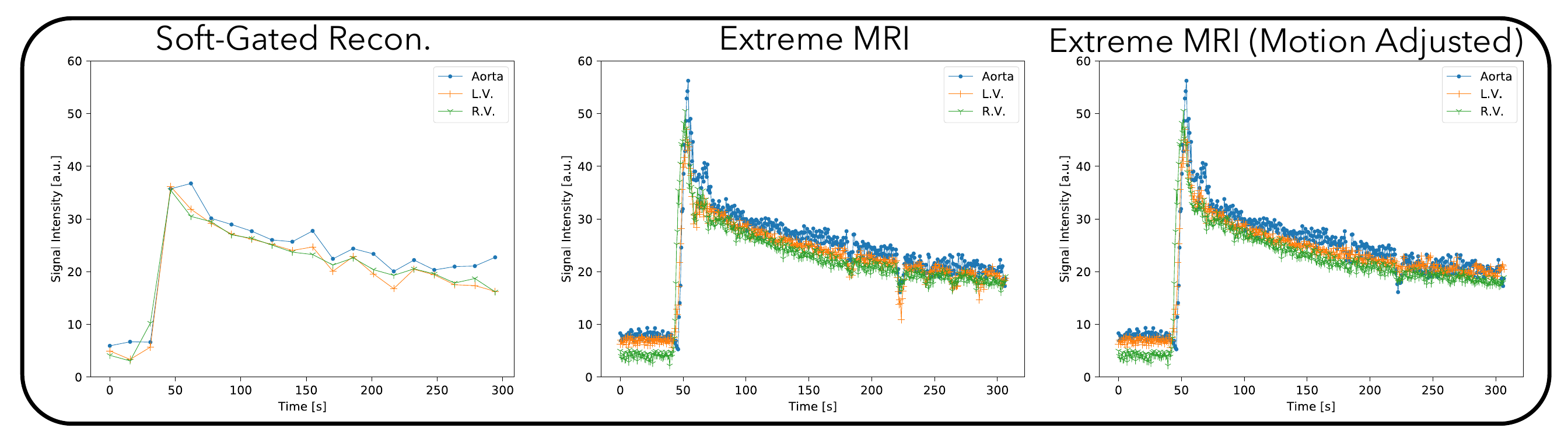}
\caption{Signal intensity curves for the proposed method and the soft-gated reconstruction of the second DCE dataset. From the signal intensity curves, the peak contrast enhancements are higher for the proposed reconstruction than for the soft-gated reconstruction. Dips in the signal intensity curves in later parts of the scan can be seen for the proposed reconstruction, which corresponds to when bulk motions occur. A motion adjusted plot was created by manually tracking the voxels over time. Variations due to bulk motions in the signal intensity plots are mostly removed after motion adjustment.}
\label{fig:dce2_curves}
\end{center}
\end{figure}

Figure~\ref{fig:dce2} and Supporting Information Video~\ref{vid:dce2_extreme} and~\ref{vid:dce2_softgated} compare the proposed method with the soft-gated reconstruction. From Supporting Information Video~\ref{vid:dce2_extreme}, the proposed reconstruction shows regular respiratory motion in the beginning, but after contrast injection, breathing becomes more rapid and the patient body shifts to the right seven times. While the image quality during these bulk movements degrades, it improves as soon as the patient body returns to the original position. Similar to the first DCE dataset, distinct phases of contrast enhancement to different organs can be seen, whereas the soft-gated reconstruction merges all dynamics into one frame, including the bulk motion. From the signal intensity curves in Figure~\ref{fig:dce2_curves}, the peak contrast enhancements are higher for the proposed reconstruction than for the soft-gated reconstruction. Dips in the signal intensity curves in later parts of the scan can be seen for the proposed reconstruction, which corresponds to when bulk motions occur. A motion adjusted plot was created by manually tracking the voxels over time. Variations due to bulk motions in the signal intensity plots are mostly removed after motion adjustment. 

An instance of the proposed method takes about 6 hours and the soft-gated reconstruction takes about 30 minutes. The resulting image using the MSLR representation takes 1GB to store.

\subsubsection{Third DCE dataset}

\begin{figure}[!ht]
\begin{center}
  \includegraphics[width=\linewidth]{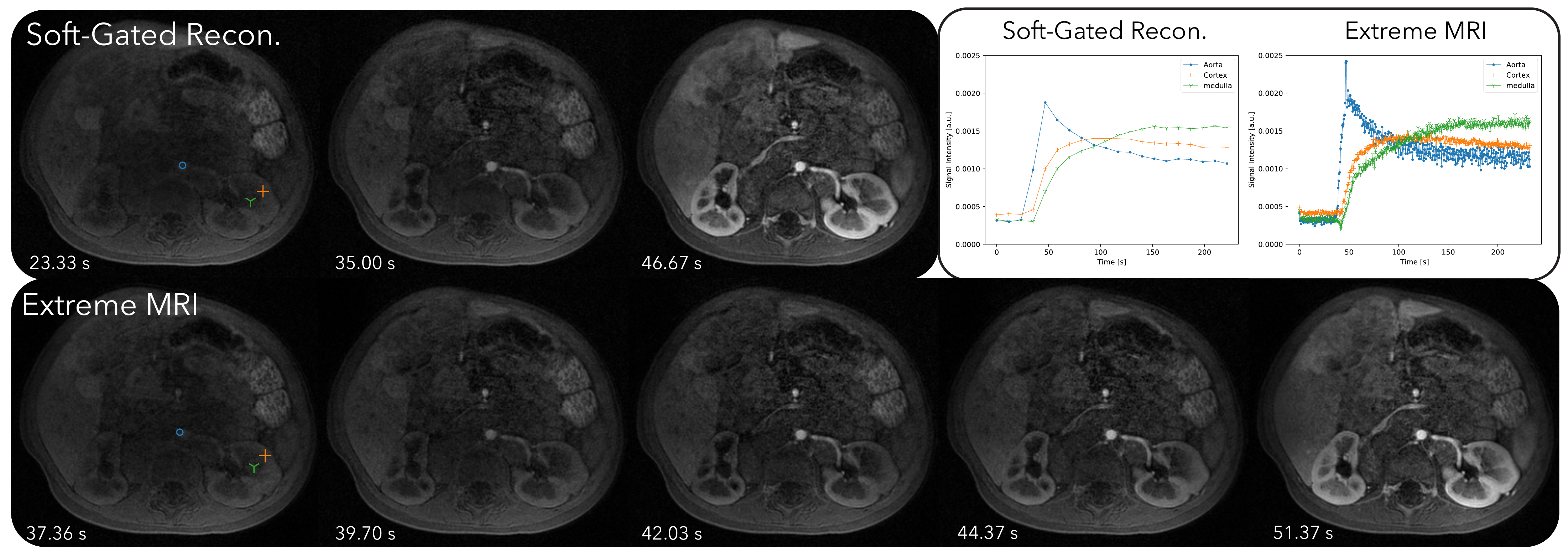}
\caption{Comparison of the proposed method with the soft-gated reconstruction of the third DCE dataset. Dynamics can be seen more clearly in Supporting Information Video~\ref{vid:dce3_extreme} and~\ref{vid:dce3_softgated}. From Supporting Information Video~\ref{vid:dce3_extreme}, regular breathing motion can be seen in the proposed reconstruction. Contrast dynamics starting from the aorta, and slowly filling in the cortex and the medulla can be observed. The aortic temporal changes can be seen at a higher frame-rate for the proposed reconstruction than for the soft-gated reconstruction.}
\label{fig:dce3}
\end{center}
\end{figure}

Figure~\ref{fig:dce3} and Supporting Information Video~\ref{vid:dce3_extreme} and~\ref{vid:dce3_softgated} compare the proposed method with the soft-gated reconstruction. From Supporting Information Video~\ref{vid:dce3_extreme}, regular breathing motion can be seen in the proposed reconstruction. Contrast dynamics starting from the aorta, and slowly filling in the cortex and the medulla can be observed. The aortic temporal changes can be seen at a higher frame-rate for the proposed reconstruction than for the soft-gated reconstruction.

An instance of the proposed method takes about 42 hours and the soft-gated reconstruction takes about 2 hours. The resulting image using the MSLR representation takes 2.5 GBs to store.

\subsection{Pulmonary Imaging with Bit-reversed Ordered 3D UTE Radial Trajectory}

\subsubsection{First lung dataset}

\begin{figure}[!ht]
\begin{center}  \includegraphics[width=0.8\linewidth]{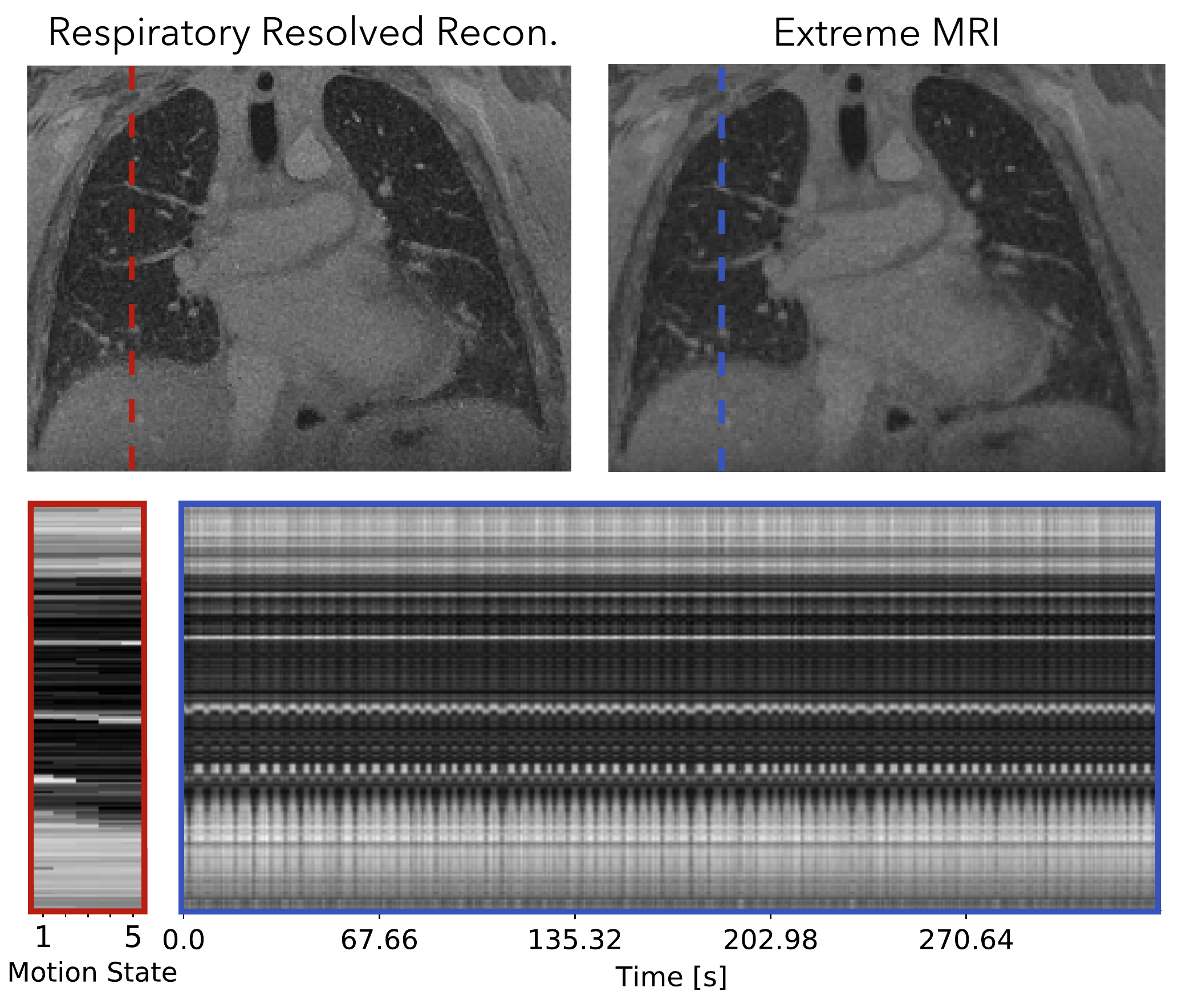}
\caption{Comparison of the proposed method with the respiratory-resolved reconstruction of the first lung dataset. Dynamics can be seen more clearly in Supporting Information Video~\ref{vid:lung1_extreme} and~\ref{vid:lung1_motionresolved}. From the cross-section over time and Supporting Information Video~\ref{vid:lung1_extreme}, regular breathing with slight variable rates can be observed. Overall, Supporting Information Video~\ref{vid:lung1_extreme} of the proposed reconstruction shows temporal flickering artifacts. Looking at each frame individually, the proposed reconstruction shows similar image quality and sharpness as the respiratory resolved reconstruction for the expiration phase. For other phases, the respiratory resolved reconstruction is slightly sharper near the diaphragms.}
\label{fig:lung1}
\end{center}
\end{figure}

Figure~\ref{fig:lung1} and Supporting Information Video~\ref{vid:lung1_extreme} and~\ref{vid:lung1_motionresolved} compare the proposed method with the respiratory-resolved reconstruction. From the cross-section over time and Supporting Information Video~\ref{vid:lung1_extreme}, regular breathing with slight variable rates can be observed. Overall, Supporting Information Video~\ref{vid:lung1_extreme} of the proposed reconstruction shows some temporal flickering artifacts. Looking at each frame individually, the proposed reconstruction shows similar image quality and sharpness as the respiratory resolved reconstruction for the expiration phase. For other phases, the respiratory resolved reconstruction is slightly sharper near the diaphragms.

An instance of the proposed method takes about 45 hours and the respiratory resolved reconstruction takes about an hour. The resulting image using the MSLR representation takes 4.4GB to store.

\subsubsection{Second lung dataset}

\begin{figure}[!ht]
\begin{center}
  \includegraphics[width=0.8\linewidth]{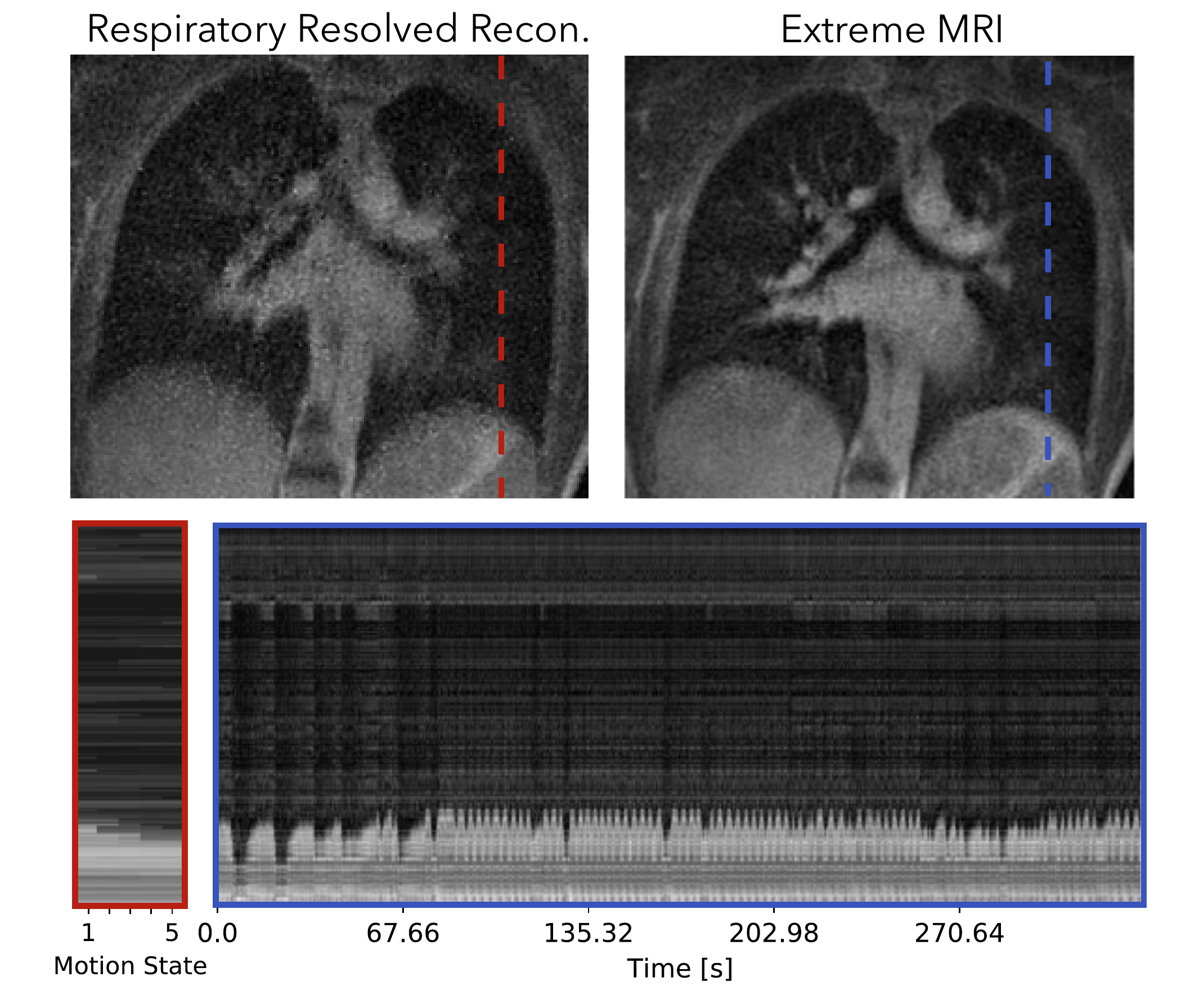}
\caption{Comparison of the proposed method with the respiratory-resolved reconstruction of the second lung dataset. Dynamics can be seen more clearly in Supporting Information Video~\ref{vid:lung2_extreme} and~\ref{vid:lung2_motionresolved}. From the cross-section and Supporting Information Video~\ref{vid:lung2_extreme}, coughing can be observed in the beginning of the scan for the proposed reconstruction. The patient can be seen to return to a more regular breathing pattern after a while but still occasionally show abrupt motions. The proposed reconstruction does show severe flickering temporal artifacts when the patient coughs, but in general has less noise-like artifacts and much sharper features than the respiratory resolved reconstruction.}
\label{fig:lung2}
\end{center}
\end{figure}

Figure~\ref{fig:lung2} and Supporting Information Video~\ref{vid:lung2_extreme} and~\ref{vid:lung2_motionresolved} compare the proposed method with the respiratory-resolved reconstruction. From the cross-section and Supporting Information Video~\ref{vid:lung2_extreme}, coughing can be observed in the beginning of the scan for the proposed reconstruction. The patient can be seen to return to a more regular breathing pattern after a while but still occasionally show abrupt motions. The proposed reconstruction does show flickering temporal artifacts when the patient coughs, but in general has less noise-like artifacts and much sharper features than the respiratory resolved reconstruction. 

An instance of the proposed method takes about 11.5 hours and the respiratory resolved reconstruction takes about 15 minutes. The resulting image using the MSLR representation takes 2.8GBs to store.

\section{Discussion}

We have developed a method to reconstruct large-scale volumetric dynamic image sequences from rapid, continuous, and non-gated acquisitions. The reconstruction problem considered is vastly underdetermined, and computationally and memory demanding. Using MSLR, the proposed technique can greatly compress dynamic image sequence on the order of a hundred GB to only a few GBs. Using the Burer-Monteiro heuristic, the reconstruction can directly optimize for the compressed representation. Finally incorporating stochastic optimization, the run-time for the resulting method is drastically reduced.

In Figure~\ref{fig:lr_compare}, we compared the effect of different LR modelings. Consistent with observations in~\cite{trzasko_local_2011, zhang_fast_2015}, LR has difficulty representing spatially localized dynamics, and exhibits temporal blurring of contrast enhancements in the ventricles. LLR is able to depict spatially localized dynamics well and show distinct contrast enhancements in the left and right ventricles. However, compared to MSLR, LLR has more spurious temporal artifacts. Because LR and LLR are subsets of MSLR, MSLR in principle can always perform better with the suitably tuned regularization parameters. The purpose of the experiment is to show that with a fixed regularization relationship between scales described in Equation~\eqref{eq:lambdas}, MSLR can still represent dynamics with the appropriate scale. This is also supported qualitatively in the MSLR decomposition shown in Figure~\ref{fig:decom}.

The benefit of using SGD compared to conventional iterative algorithms can be seen in Figure~\ref{fig:convergence}. For GD, each pass of the k-space data only computes one iteration, whereas for SGD, each pass computes $TC$ many iterations. Even though each SGD iteration makes fewer progress than GD, SGD reaches an approximate minimum faster by performing many more iterations. In our experiment with 20 frames, we observe that GD is approximately 15x slower, and expect this difference to be even greater with 500 frames. However, the current SGD reconstruction time is still long, ranging from 6 hours to 42 hours, depending on the number of coils and the reconstruction matrix size. We remain hopeful that with additional computing devices, either with large number of GPUs or cloud computing, reconstruction time can be brought to a reasonable range. This is partly supported by Figure~\ref{fig:convergence}b, which shows that the speedup of using multiple GPUs is almost linear from one to four GPUs.

Figure~\ref{fig:dce1},~\ref{fig:dce2}, and~\ref{fig:dce3} all show that the proposed reconstruction displays much finer dynamics that are not represented in soft-gated reconstructions with low frame-rates. Distinct phases of contrast enhancements in different organs can be seen, which are more physiologically accurate. The benefits of higher temporal resolution can also be seen from signal intensity curves. In particular, signal intensity peaks of the aorta are much higher in the proposed reconstruction, but are averaged out in the soft-gated reconstruction. While bulk motion still affects the overall image quality as shown in Figure~\ref{fig:dce2}, the proposed reconstruction allows us to retrospectively adjust for bulk motion when computing the signal intensity curves, which can be useful for quantification purposes.

Figure~\ref{fig:lung2} shows that for pulmonary imaging, the proposed reconstruction really shines when there are non-periodic motions. Because Figure~\ref{fig:lung1} mostly consists regular breathing, the proposed reconstruction does not offer much beyond unrolling the periodic dynamics. However, for the second dataset with irregular breathing, the proposed reconstruction provides substantially improved image quality, and also depicts the irregularities. In particular, coughing can be seen from the beginning of the dynamic image series. 

While the proposed method does show transient dynamics that are not seen in gating based reconstructions, flickering temporal artifacts can often be seen in reconstructions. \revised{\rnum{1.1}We believe the main cause for the flickering artifacts is that block low rank models are inherently sensitive to local fluctuations. Each block is almost independent with others, and small blocks are more easily fitted to aliasing artifacts and noise. This explains why LLR has flickering and LR does not. Ideally, MSLR would suppress the spurious blocks and use larger blocks instead. However, in the limited memory setting, we can only use three scales, and the model is forced to use smaller blocks for representation. This is why for larger datasets (the third DCE dataset and all lung datasets), we used larger block sizes for reconstruction. If more memory is available, we would do more scales and let the reconstruction choose the appropriate blocks.}


Besides the flickering artifacts, the proposed technique also shows degraded image quality and temporal blurring for large bulk motions, such as those in Supporting Information Video~\ref{vid:dce2_extreme}. One reason is that MSLR, or in general LR representations, does not explicitly model for motions. Hence, for large displacements, the resulting dynamics may not be low rank even for small block sizes. While there are existing works~\cite{otazo_motion-guided_2014} on incorporating motion information into LR modelling, it is still unclear how to do it in a computation and memory friendly way for the extreme MRI setting.

\revised{\rnum{2.3}We would also like to emphasize that the actual temporal resolution of our reconstruction is lower than the prescribed temporal resolution. This is especially true with the finite-difference temporal penalty applied on the temporal bases. Since we do not have access to the ground truths, we are unable to quantify the actual spatiotemporal resolution. A potential future direction is to create a realistic large-scale digital phantom to help quantify the actual resolution.}

\revised{\rnum{1.4}In terms of clinical applications, the proposed method is suitable when volumetric information with high temporal resolution is desired. In particular, it could potentially work well in the lower abdomen where conventional techniques are often affected by bowel motion. Also, functional imaging approaches that require large volume coverage, such as imaging of tracheomalacia, could benefit from the technique. Another potential application is MR spirometry~\cite{voorhees_magnetic_2005} for irregular breathing, in which regional lung deformations over times are used to estimate pulmonary function. If the reconstruction time can be made short enough with more advanced hardware, the proposed method can also be useful for interventional MRI.}

\revised{\rnum{2.1}Finally, we would like to mention several theoretical results regarding the Burer-Monteiro factorization and point to potential directions to bridge the gaps between their assumptions and our setting. Existing works~\cite{burer_nonlinear_2003, journee_low-rank_2010} have shown that any second-order stationary point (i.e. solutions with zero gradient and positive semi-definite Hessian) of the low rank factorized non-convex problem is equivalent to the global minimum of the convex problem using nuclear norm, as long as the solution matrix rank is smaller than what is prescribed. Recent theoretical results~\cite{zhu_global_2018} have further shown that under idealized incoherence conditions (such as the restricted isometry property), any second-order stationary point of the non-convex formulation is also a global minimum. In order to generalize these results to our setup, it is crucial to extend inherent conditions to multi-channel forward models. In addition, it remains to be shown that 3D radial and 3D cones trajectories with pseudo-randomized orderings satisfy these theoretical assumptions.}

\section{Conclusion}

We demonstrated a method that can reconstruct massive 3D dynamic image series in the extreme undersampling and extreme computation setting. The proposed technique shows transient dynamics that are not seen in gating based methods. When applied to datasets with irregular,  or non-repetitive motions, the proposed method also displays sharper image features.

\bibliography{main}

\clearpage

\section*{List of Supporting Information Videos}

\begin{enumerate}[label=S\arabic*]
\item 3D rendering of the proposed reconstruction of the first DCE dataset. The proposed method can reconstruct large-scale volumetric dynamic MRI, which enables high quality reformatting as shown. \label{vid:dce1_3drender}
\item The proposed reconstruction with LR modeling of the first DCE dataset. LR shows contrast flowing into left and right ventricles at the same time, which is not physiological correct.\label{vid:dce1_lr}

\item The proposed reconstruction with LLR modeling of the first DCE dataset. LLR shows contrast enhancing first in the right ventricle then the left ventricle, which is more physiological accurate, but displays more flickering temporal artifacts.\label{vid:dce1_llr}

\item The proposed reconstruction with MSLR modeling of the first DCE dataset. MSLR achieves a balance between representing contrast enhancement dynamics and reducing artifacts. Contrast enhancements can be seen starting from the right ventricle, to the lung, then the left ventricle, and to the aorta.\label{vid:dce1_extreme}

\item MSLR decomposition of the first DCE dataset. The scale with the 128$^3$-sized blocks mostly shows static background tissues. The scale with the 64$^3$-sized blocks depicts mostly contrast enhancements in the heart and aorta, and respiratory motion. The scale with 32$^3$-sized blocks displays spatially localized dynamics, such as those in the right ventricle, and oscillates more over time than other scales.\label{vid:dce1_mslr_decom}

\item Soft-gated reconstruction of the first DCE dataset. The soft-gated reconstruction merges together the temporal changes of the lung, the left ventricle and the aorta.\label{vid:dce1_softgated}

\item The proposed reconstruction of the second DCE dataset. The proposed reconstruction shows regular respiratory motion in the beginning, but after contrast injection, breathing becomes more rapid and the patient body shifts to the right seven times. While the image quality during these bulk movements degrades, it improves as soon as the patient body returns to the original position.\label{vid:dce2_extreme}

\item Soft-gated reconstruction of the second DCE dataset. The soft-gated reconstruction merges all dynamics during contrast injection into one frame, including the bulk motion. \label{vid:dce2_softgated}

\item The proposed reconstruction of the third DCE dataset. Regular breathing motion can be seen in the proposed reconstruction. Contrast dynamics starting from the aorta, and slowly filling in the cortex and the medulla can be observed at a high frame-rate.\label{vid:dce3_extreme}

\item Soft-gated reconstruction of the third DCE dataset. Contrast dynamics starting from the aorta, and slowly filling in the cortex and the medulla can be observed. \label{vid:dce3_softgated}

\item The proposed reconstruction of the first lung dataset. Regular breathing with slight variable rates can be observed. Overall, the proposed reconstruction shows temporal flickering artifacts. \label{vid:lung1_extreme}

\item Respiratory resolved reconstruction of the first lung dataset. The respiratory resolved reconstruction is slightly sharper near the diaphragms in most phases than the proposed reconstruction.\label{vid:lung1_motionresolved}

\item The proposed reconstruction of the second lung dataset. Coughing can be observed in the beginning of the scan for the proposed reconstruction. The patient can be seen to return to a more regular breathing pattern after a while but still occasionally show abrupt motions. The proposed reconstruction does show severe flickering temporal artifacts during the patient coughing, but in general has less noise-like artifacts and much sharper features than the respiratory resolved reconstruction.\label{vid:lung2_extreme}

\item Respiratory resolved reconstruction of the second lung dataset. The respiratory resolved reconstruction in general more noise-like artifacts and much blurrier features than the proposed method.\label{vid:lung2_motionresolved}

\end{enumerate}

\clearpage



\end{document}